\documentclass[preprint,preprintnumbers,showpacs,amssymb,amsmath,secnumarabic,tightenlines,nofootinbib,12pt]{revtex4}
\usepackage{bm}
\usepackage{psfig}
\def\babar{\mbox{\sl B\hspace{-0.4em} {\scriptsize\sl A}\hspace{-0.4em} 
	B\hspace{-0.4em} {\scriptsize\sl A\hspace{-0.1em}R}}}
\def\fom{figure of merit}
\def\spr{StatPatternRecognition}

\begin{document}

\preprint{}

\title{Optimization of Signal Significance by Bagging Decision Trees
\footnote{Work partially supported by Department of Energy under Grant
DE-FG03-92-ER40701.}
}
\author{Ilya Narsky}
\email{narsky@hep.caltech.edu}
\affiliation{California Institute of Technology}
\date{\today}

\begin{abstract}
An algorithm for optimization of signal significance or any other
classification \fom\ suited for analysis of high energy physics (HEP)
data is described. This algorithm trains decision trees on many
bootstrap replicas of training data with each tree required to
optimize the signal significance or any other chosen \fom. New data
are then classified by a simple majority vote of the built trees. The
performance of this algorithm has been studied using a search for the
radiative leptonic decay $B\to\gamma l\nu$ at \babar\ and shown to be
superior to that of all other attempted classifiers including such
powerful methods as boosted decision trees. In the $B\to\gamma e\nu$
channel, the described algorithm increases the expected signal
significance from $2.4\sigma$ obtained by an original method designed
for the $B\to\gamma l\nu$ analysis to $3.0\sigma$.
\end{abstract}


\maketitle

\section{Introduction}

Separation of signal and background is perhaps the most important
problem in analysis of HEP data. Various pattern classification tools
have been employed in HEP practice to solve this problem. Fisher
discriminant~\cite{fisher} and feedforward backpropagation neural
networks~\cite{nn} are the two most popular methods chosen by HEP
analysts at present. Alternative algorithms for classification such as
decision trees~\cite{cart}, bump hunting~\cite{prim}, and
AdaBoost~\cite{boosted} have been recently explored by the HEP
community as well~\cite{modern_hep,roe,spr}. These classifiers can be
characterized by such features as predictive power, interpretability,
stability and ease of training, CPU time required for training and
classifying new events, and others. It is important to remember that
the choice of a classifier for each problem should be driven by
specifics of the analysis. For example, if the major goal of pattern
classification is to achieve a high quality of signal and background
separation, flexible classifiers such as AdaBoost and neural nets
should be the prime choice. While neural nets generally perform quite
well in low dimensions, they become too slow and unstable in
high-dimensional problems losing the competition to AdaBoost. If the
analyst, however, is mostly concerned with a clear interpretation of
the classifier output, decision trees and bump hunting algorithms are
a more appealing option. These classifiers produce rectangular
regions, easy to visualize in many dimensions.

One of the problems faced by HEP analysts is the indirect nature of
available classifiers. In HEP analysis, one typically wants to
optimize a \fom\ expressed as a function of signal and background, $S$
and $B$, expected in the signal region. An example of such \fom\ is
signal significance, $S/\sqrt{S+B}$, often used by physicists to
express the cleanliness of the signal in the presence of statistical
fluctuations of observed signal and background. None of the available
popular classifiers optimizes this \fom\ directly. CART~\cite{cart}, a
popular commercial implementation of decision trees, splits training
data into signal- and background-dominated rectangular regions using
the Gini index, $Q=2p(1-p)$, as the optimization criterion, where $p$
is the correctly classified fraction of events in a tree node. Neural
networks typically minimize a quadratic classification error,
${\mathcal E}_{\mbox{qua}} = \sum_{n=1}^N (y_n-f(x_n))^2$, where $y_n$
is the true class of an event, -1 for background and 1 for signal,
$f(x_n)$ is the continuous value of the class label in the range
$[-1,1]$ predicted by the neural network, and the sum is taken over
$N$ events in the training data set. Similarly, AdaBoost minimizes an
exponential classification error, ${\mathcal E}_{\mbox{exp}}
=\sum_{n=1}^N \exp(-y_n f(x_n))$. These optimization criteria are not
necessarily optimal for maximization of the signal significance. The
usual solution is to build a neural net or an AdaBoost-based
classifier and then find an optimal cut on the continuous output of
the classifier to maximize the signal significance. For decision
trees, the solution is to construct a decision tree with many terminal
nodes and then combine these nodes to maximize the signal
significance.

This problem has been partially addressed in my C++ software package
for pattern classification~\cite{spr}. Default implementations of the
decision tree and the bump hunting algorithm include both standard
figures of merit used for commercial decision trees such as the Gini
index and HEP-specific figures of merit such as the signal
significance or the signal purity, $S/(S+B)$. The analyst can optimize
an arbitrary \fom\ by providing an implementation to the corresponding
abstract interface set up in the package.

AdaBoost and the neural net, however, cannot be modified that
easily. The functional forms of the classification error are
intimately tied to implementations of these two classification
algorithms. Finding a powerful method for optimization of HEP-specific
figures of merit is therefore an open question.

This note describes an algorithm that can be used for direct
optimization of an arbitrary \fom. Optimization of the signal
significance by this algorithm has shown results comparable or better
than those obtained with AdaBoost or the neural net. The training time
used by this algorithm is comparable to that used by AdaBoost with
decision trees; the algorithm is therefore faster than the neural net
in high dimensions. The method has been coded in C++ and included in
the \spr\ package available for free distribution to HEP analysts.

\section{Bagging Decision Trees}

The implementation of decision trees used for the proposed algorithm
is described in detail in Ref.~\cite{spr}. The key feature of this
implementation is its ability to optimize HEP-specific figures of
merit such as the signal significance.

A decision tree, even if it directly optimizes the desired \fom, is
rarely powerful enough to achieve a good separation between signal and
background. The tree produces a set of signal-dominated rectangular
regions. Rectangular regions, however, often fail to capture a
non-linear structure of data. The mediocre predictive power of a
single decision tree can be greatly enhanced by one of the two popular
methods for combining classifiers --- boosting and bagging.

Both these methods work by training many classifiers, e.g., decision
trees, on variants of the original training data set. A boosting
algorithm enhances weights of misclassified events and reduces weights
of correctly classified events and trains a new classifier on the
reweighted sample. The output of the new classifier is then used to
re-evaluate fractions of correctly classified and misclassified events
and update the event weights accordingly. After training is completed,
events are classified by a weighted vote of the trained classifiers.
AdaBoost, a popular version of this approach, has been shown to
produce a high-quality robust training mechanism. Application of
AdaBoost to HEP data has been explored in Refs.~\cite{roe,spr}.

In contrast, bagging algorithms~\cite{bagging} do not reweight
events. Instead, they train new classifiers on bootstrap replicas of
the training set. Each bootstrap replica~\cite{bootstrap} is obtained
by sampling with replacement from the original training set, with the
size of each replica equal to that of the original set. After training
is completed, events are classified by the majority vote of the
trained classifiers. For successful application of the bagging
algorithm, the underlying classifier must be sensitive to small
changes in the training data. Otherwise all trained classifiers will
be similar, and the performance of the single classifier will not be
improved. This condition is satisfied by a decision tree with fine
terminal nodes. Because of the small node size each decision tree is
significantly overtrained; if the tree were used just by itself, its
predictive power on a test data set would be quite poor. However,
because the final decision is made by the majority vote of all the
trees, the algorithm delivers a high predictive power.

Various kinds of boosting and bagging algorithms have been compared in
the statistics literature. Neither of these two approaches has a clear
advantage over the other. On average, boosting seems to provide a
better predictive power. Bagging tends to perform better in the
presence of outliers and significant noise~\cite{comparison}.

For optimization of the signal significance, however, bagging is the
choice favored by intuition. Reweighting events has an unclear impact
on the effectiveness of the optimization routine with respect to the
chosen \fom. While it may be possible to design a reweighting
algorithm efficient for optimization of a specific \fom, at present
such reweighting algorithms are not known. Bagging, on the other hand,
offers an obvious solution. If the base classifier directly optimizes
the chosen \fom, bagging is equivalent to optimization of this \fom\
integrated over bootstrap replicas. In effect, the bagging algorithm
finds a region in the space of physical variables that optimizes the
expected value of the chosen \fom\ --- exactly what the analyst is
looking for.

Bagging decision trees is certainly not a new item in the statistics
research. The only novelty introduced in this note is the decision
tree designed for direct optimization of an arbitrary \fom, e.g., the
signal significance.

\begin{figure}[htbp]
   \begin{center}
   \vskip -2.0 cm
   \hbox{
   \quad
   \parbox[t]{16.5cm}{ \psfig{file=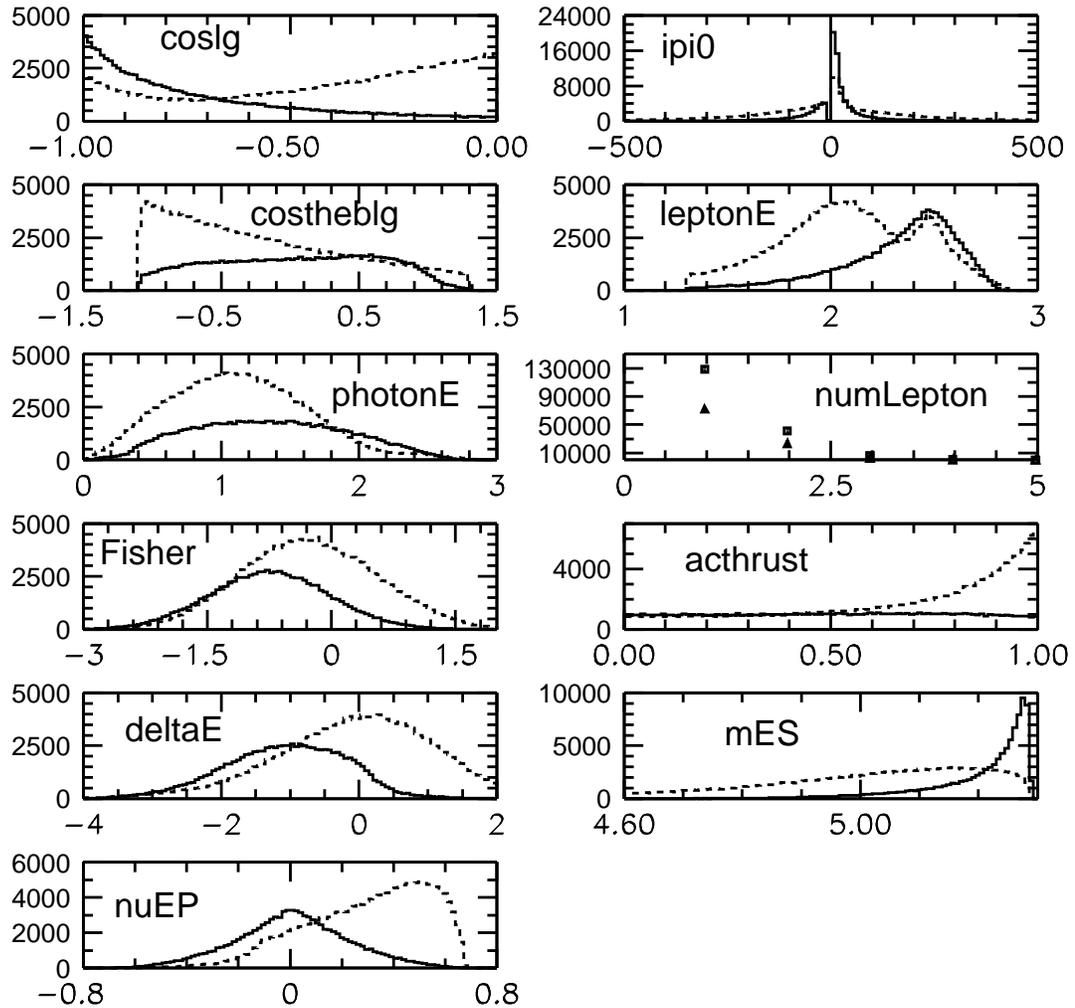,width=16.5cm}
   \caption[c]{\small  \label{fig:LNuGamma_e_Vars}  
	Separation variables for the $B\to\gamma l\nu$ analysis. 
	Signal MC is shown with a solid line 
	(triangles in the {\tt numLepton} plot), 
	and the overall combined background is shown with a dashed line
	(squares in the {\tt numLepton} plot).
	} }
   }
   \quad
   \end{center}
\end{figure}

\begin{table}[bthp]
\begin{center}
\label{tab:lnugamma_signif}
\caption{Signal significance, ${\mathcal S}_{\mbox{train}}$,
${\mathcal S}_{\mbox{valid}}$, and ${\mathcal S}_{\mbox{test}}$, for
the $B\to\gamma l\nu$ training, validation, and test samples obtained
with various classification methods. The signal significance computed
for the test sample should be used to judge the predictive power of
the included classifiers.  A branching fraction of $3\times 10^{-6}$
was assumed for both $B\to\gamma \mu\nu$ and $B\to\gamma e\nu$
decays. $W_1$ and $W_0$ represent the signal and background,
respectively, expected in the signal region after the classification
criteria have been applied; these two numbers have been estimated
using the test samples. All numbers have been normalized to the
integrated luminosity of $210\ \mbox{fb}^{-1}$. The best value of the
expected signal significance is shown in boldface.}
\begin{tabular}{|c|c|c|c|c|c|c|c|c|c|c|}\hline
Method 
& \multicolumn{5}{c|}{$B\to\gamma e\nu$} 
& \multicolumn{5}{c|}{$B\to\gamma \mu\nu$} \\ \cline{2-11}
& ${\mathcal S}_{\mbox{train}}$
& ${\mathcal S}_{\mbox{valid}}$ & ${\mathcal S}_{\mbox{test}}$ & $W_1$ & $W_0$
& ${\mathcal S}_{\mbox{train}}$
& ${\mathcal S}_{\mbox{valid}}$ & ${\mathcal S}_{\mbox{test}}$ & $W_1$ & $W_0$
\\ \hline

Original method 
& 2.66 & - & 2.42 & 37.5 & 202.2 
& 1.75 & - & 1.62 & 25.8 & 227.4 \\ \hline

Decision tree
& 3.28 & 2.72 & 2.16 & 20.3 & 68.1 
& 1.74 & 1.63 & 1.54 & 29.0 & 325.9 \\ \hline

Bump hunter with one bump 
& 2.72 & 2.54 & 2.31 & 47.5 & 376.6
& 1.76 & 1.54 & 1.54 & 31.7 & 393.8 \\ \hline

AdaBoost with binary splits 
& 2.53 & 2.65 & 2.25 & 76.4 & 1077.3 
& 1.66 & 1.71 & 1.44 & 45.2 & 935.6 \\ \hline

AdaBoost with decision trees
& 13.63 & 2.99 & 2.62 & 58.0 & 432.8
& 11.87 & 1.97 & 1.75 & 41.6 & 523.0 \\ \hline

Combiner of background subclassifiers
& 3.03 & 2.88 & 2.49 & 83.2 & 1037.2 
& 1.84 & 1.90 & 1.66 & 55.2 & 1057.1 \\ \hline

Bagging decision trees
& 9.20 & 3.25 & {\bf 2.99} & 69.1 & 465.8
& 8.09 & 2.07 & {\bf 1.98} & 49.4 & 571.1 \\ \hline

\end{tabular}
\end{center}
\end{table}

\begin{figure}[htbp]
   \begin{center}
   \hbox{\hskip -0cm
   \quad 
   \parbox[t]{5.5cm}{ 
     \psfig{figure=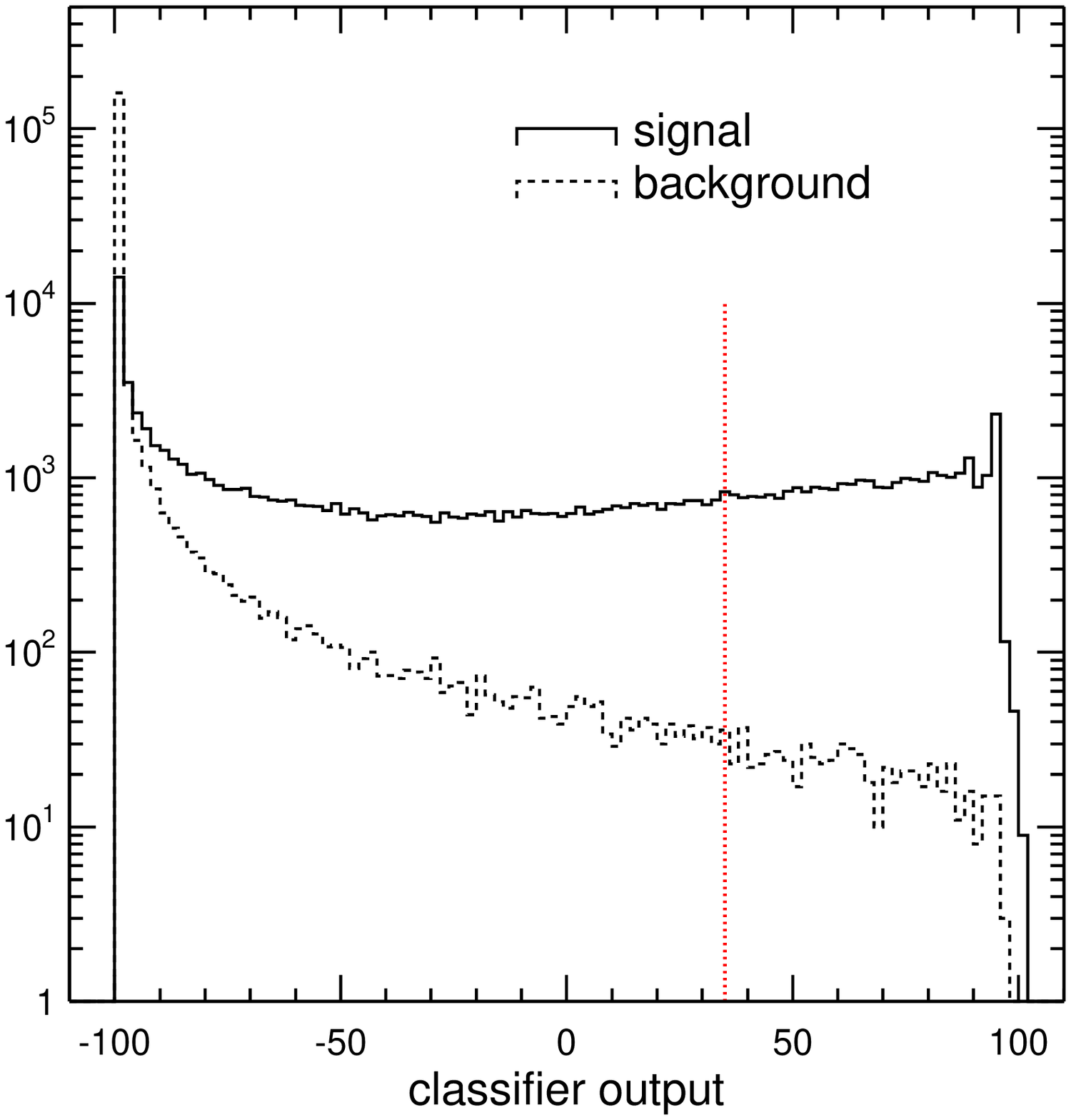,width=8cm}
   } 
   \quad
   \parbox[t]{5.5cm}{ 
     \psfig{figure=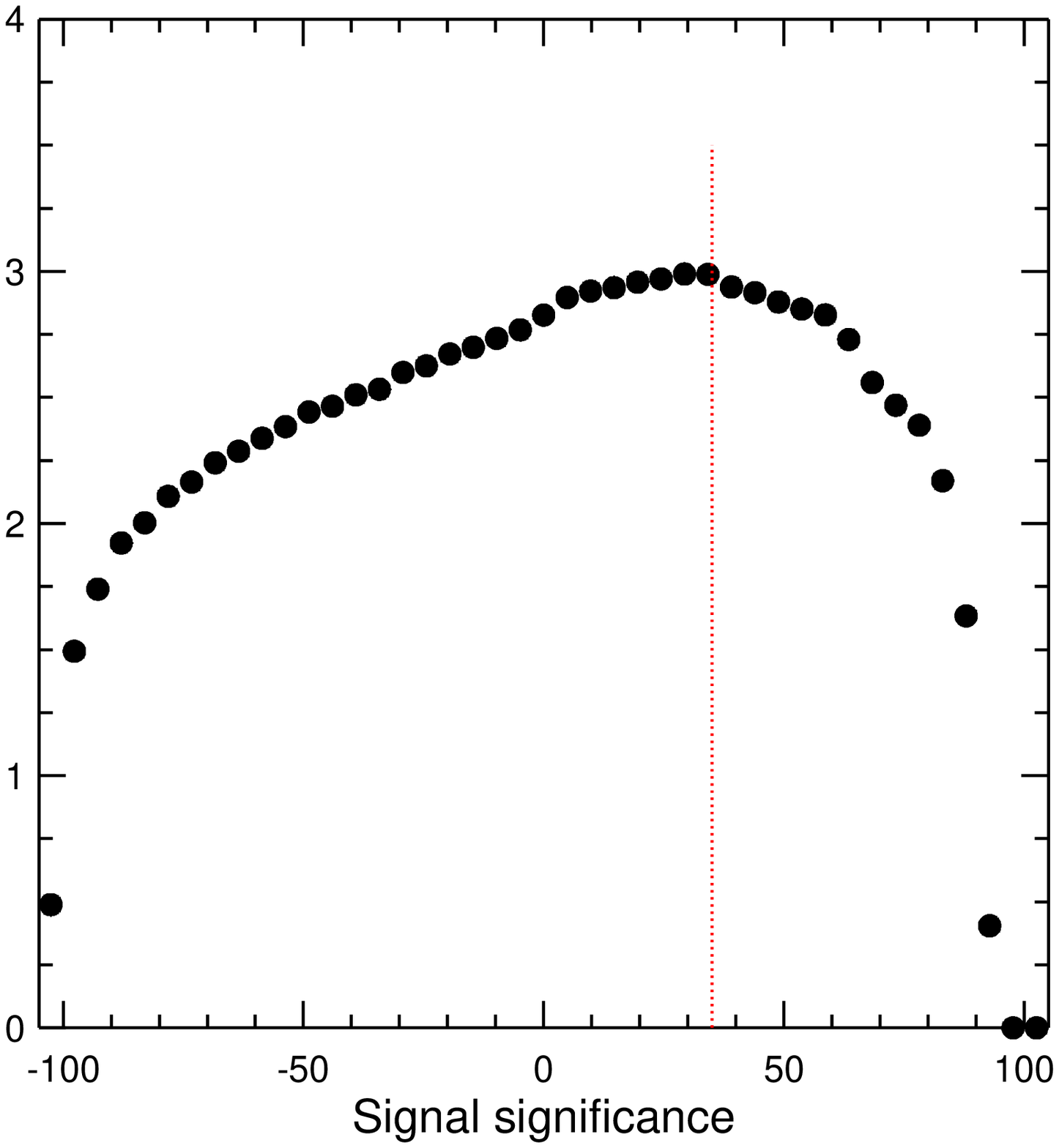,width=8cm} }
   }
   \caption[c]{\small \label{fig:Bagger_LeGamma} 
	Output of the bagging algorithm 
	with 100 trained decision trees (left) 
	and the signal significance
	versus the lower cut on the output (right)
	for the $B\to\gamma e\nu$ test sample.
	The cut maximizing the signal significance,
	obtained using the validation sample, 
	is shown with a vertical line.
	}
   \end{center} 
\end{figure}

Performance of the described bagging algorithm has been studied using
a search for the radiative leptonic decay $B\to\gamma l\nu$ at
\babar. Eleven variables used for classification in this analysis are 
shown in Fig.~\ref{fig:LNuGamma_e_Vars}. Several methods have been
used to separate signal from background by maximizing the signal
significance: an original method developed by the analysts, the
decision tree optimizing the signal significance, the bump hunting
algorithm, AdaBoost with binary splits, AdaBoost with decision trees
optimizing the Gini index, and an AdaBoost-based combiner of
background subclassifiers. I also attempted to use a feedforward
backpropagation neural network with one hidden layer, but the network
was unstable and it failed to converge to an optimum. A more detailed
description of this analysis and used classifiers can be found in
Ref.~\cite{spr}.

To test the bagging algorithm described in this note, I trained 100
decision trees on bootstrap replicas of the training data. For
classification of new data, the trained trees were combined using an
algebraic sum of their outputs: if an event was accepted by a tree,
the output for this event was incremented by 1 and decremented by 1
otherwise.  The minimal size of the terminal node in each tree, 100
events for both $B\to\gamma e\nu$ and $B\to\gamma\mu\nu$ channels, was
chosen by comparing values of the signal significance computed for the
validation data. The size of the trained decision trees varied from
390 to 470 terminal signal nodes in the $B\to\gamma e\nu$ channel and
from 300 to 370 in the $B\to\gamma\mu\nu$ channel. Jobs executing the
algorithm took several hours in a batch queue at SLAC. To assess the
true performance of the method, the signal significance was then
evaluated for the test data.

All attempted classifiers are compared in Table~I. The output of the
described bagging algorithm for the $B\to\gamma e\nu$ test data is
shown in Fig.~\ref{fig:Bagger_LeGamma}. The bagging algorithm provides
the best value of the signal significance. It gives a $24\%$
improvement over the original method developed by the analysts and
shown in the first line of Table~I, and a $14\%$ improvement over
AdaBoost with decision trees shown in line 5 of Table~I; both numbers
are quoted for the $B\to\gamma e\nu$ channel.

I also used AdaBoost with decision trees optimizing the signal
significance and the bagging algorithm with decision trees optimizing
the Gini index. The first method performed quite poorly; the signal
significance obtained with this method was much worse than that
obtained by AdaBoost with decision trees optimizing the Gini
index. The bagging algorithm with decision trees optimizing the Gini
index showed an $8\%$ improvement in the $B\to\gamma e\nu$ signal
significance compared to AdaBoost with decision trees optimizing the
Gini index. But the signal significance obtained with this method was
$9\%$ worse than that obtained by the bagging algorithm with decision
trees optimizing the signal significance. The $14\%$ improvement of
the proposed bagging algorithm over AdaBoost with decision trees
originated therefore from two sources:
\begin{itemize}
	\item Using bagging instead of boosting.  

	\item Using the signal significance instead of the Gini index
	as the \fom\ for the decision tree optimization.
\end{itemize}

In an attempt to improve the signal significance even further, I used
the random forest approach~\cite{forest}, a more generic resampling
method. In addition to generating a new bootstrap replica for each
tree, I resampled the data variables used to split each node of the
tree. Because a bootstrap replica contains on average $63\%$ of
distinct entries from the original set, only 6.9 variables out of 11
were used on average to split the tree nodes. This approach showed
only a minor $1\%$ improvement in the $B\to\gamma e\nu$ signal
significance over the bagging algorithm without variable resampling.

As shown in Fig.~\ref{fig:Bagger_LeGamma}, the described bagging
algorithm does not provide a good separation between signal and
background in terms of the quadratic or exponential classification
error. It misclassifies a large fraction of signal events. However,
the method does the job it was expected to do --- it finds a region in
the space of physical variables that, on average, maximizes the signal
significance.

\section{Summary}

A bagging algorithm suitable for optimization of an arbitrary \fom\
has been described. This algorithm has been shown to give a significant
improvement of the signal significance in the search for the radiative
leptonic decay $B\to\gamma l\nu$ at \babar. Included in the \spr\
package~\cite{spr}, this method is available to HEP analysts.

\begin{acknowledgments}
Thanks to Frank Porter for comments on a draft of this note.
\end{acknowledgments}


\end{document}